\documentclass[12pt]{article}

\usepackage[vcentermath]{youngtab}
\usepackage{cite}

\title{On massive spin-2 in the Fradkin-Vasiliev formalism.\\
II. General massive case.}

\author{M.V. Khabarov\thanks{maksim.khabarov@ihep.ru},
Yu.M. Zinoviev\thanks{Yurii.Zinoviev@ihep.ru}
\\[0.5cm]
\it{\small Institute for High Energy Physics of National
Research Center "Kurchatov Institute"} \\
\it{\small Protvino, Moscow Region, 142281, Russia}}

\date{}

\begin{document}

\maketitle

\begin{abstract}
In this work we apply the Fradkin-Vasiliev formalism based on the
frame-like gauge invariant description of the massive and massless
spin 2 to the construction of the cubic interactions vertices for
massive spin 2 self-interaction as well as its gravitational
interaction. In the first case we show that the vertex can be reduced
(by field redefinitions) to the set of the trivially gauge invariant
terms. There are four such terms which are not equivalent om-shell and
do not contain more than four derivatives. Moreover, one their
particular combination reproduces the minimal (with no more than two
derivatives) vertex. As for the gravitational vertex, we show that due
to the presence of the massless spin 2  there exist two abelian
vertices (besides the three trivially gauge invariant ones) which are
not equivalent to any trivially gauge invariant terms and can not be
removed by field redefinitions. Moreover, their existence appears to
be crucial for the possibility to reproduce the minimal two
derivatives vertex.
\end{abstract}

\thispagestyle{empty}
\newpage
\setcounter{page}{1}

\section{Introduction}

Gauge invariance serves as a main guiding principle for the
investigation of the consistent interactions for the massless higher
spin fields. It severely restricts possible interactions and provides
their complete classification. The similar problem for the massive
higher spin fields appears to be much more difficult. The well known
description of the free massive higher spins \cite{SH74,SH74a} does
not have any gauge symmetry and requires an introduction of a lot of
auxiliary fields. The free Lagrangian is constructed in such a way
that all  constrains, which are necessary to exclude all auxiliary and
nonphysical degrees of freedom, follow from the Lagrangian equations. 
In principle, one can try to construct an interacting Lagrangian such
that all the required constrains still follow, but even for the
massive spin 2 case such approach appears to be rather complicated
(see e.g. \cite{BKP99,BGKP99,Zin13}). Till now the most general
classification of the cubic interaction vertices for arbitrary spins
massless and/or massive fields has been developed in the so-called
light-cone formalism \cite{Met05,Met07b}. As a Lorentz covariant
analogue of the light-cone formalism one can use a so-called
TT-approach \cite{JLT12,JLT12a,JTW13}, where one works with the fields
which already satisfy the transversality and tracelessnes conditions.
For the massive case (where there is no any gauge symmetry) any
vertex, which can be constructed is considered to be the correct one.
In this, one assumes that it may be possible to relax the
TT-conditions and restore all necessary auxiliary fields, but, as  far
as we know, it has never been shown.

Taking into account a crucial role, that gauge invariance plays for
the massless higher spins, it seems natural to extend this notion to
the massive higher spins as well. It is indeed appears to be possible
due to the introduction of the so-called Stueckelberg fields. This has
been shown in a number of different approaches such as metric-like
\cite{Zin01,Met06}\footnote{Note that the classic results of
\cite{SH74,SH74a} can be reproduced by gauge fixing; moreover, it is
this procedure that allowed to generalize these Lagrangians into
(anti) de Sitter background.}, BRST approach
\cite{BK05,BKRT06,BKL06,BKR07,BKT07} (see \cite{FT08} for review),
the quartet unconstrained formalism \cite{BG08} and the
frame-like gauge invariant one
\cite{Zin08b,Zin08c,Zin09b,Zin09c,PV10,KhZ19}. Some applications of
such approach to the construction of the interaction vertices have
already appeared, e.g. in the metric-like \cite{Zin06,Zin09,BDGT18}, 
and the frame-like \cite{Zin10,Zin10a,Zin11,Zin14,BSZ14,Zin18,KhZ21}
formalisms. The most close to the light-cone classification results
were obtained in the BRST-BV formalism \cite{Met12} in terms of the
reducible sets of fields (see, however, \cite{BR21}).

In  any investigation of the interaction vertices one has to take
into account possible field redefinitions. In many such cases one
often restrict oneself with the redefinitions which do not raise the
number of derivatives in the vertex. What happens if one relax all
such restrictions, working with the Stueckelberg description for the
massive fields, has been investigated recently in \cite{BDGT18}. At
first, it has been shown that there always exist enough field
redefinitions to bring the vertex into an abelian form. Note, that we
call the vertex abelian if its gauge invariance requires some
non-trivial corrections to the gauge transformations but they are such
that the commutator of the gauge transformations is zero and the
algebra remains to be abelian. Moreover, by using further (even higher
derivative) field redefinitions, any vertex can be rewritten in the
trivially gauge invariant form, i.e. in terms of the gauge invariant
objects of the free theory.

Recently \cite{KhZ21} we considered this problem in the frame-like
gauge invariant formalism using a gravitational interaction for
massive spin 3/2 as an example. From one hand, we have shown that in
this case it is also possible to convert the vertex into the abelian
form by the appropriate field redefinitions. From the other hand, it
appeared that due to the presence of massless graviton there exists a
couple of abelian vertices which are not equivalent to any trivially
gauge invariant ones and can not be removed by field redefinitions.
Moreover, the existence of these abelian vertices was crucial for the
possibility to reproduce the minimal (with no more than one
derivative) gravitational vertex.

In this paper working in the Fradkin-Vasiliev formalism
\cite{FV87,FV87a,Vas11} based on the frame-like description of massive
and massless spin 2 \cite{Zin08b,PV10}, we consider cubic vertices for
massive spin 2 self-interactions as well as its gravitational
interaction. Let us briefly collect our findings here. \\
{\bf Self-interaction}
\begin{itemize}
\item We considered the most general non-abelian ansatz for such
vertex and showed that it indeed can be converted into the abelian
form by field redefinitions.
\item We have shown that all the abelian vertices are equivalent to
some trivially gauge invariant ones and/or can be removed by field
redefinitions.
\item There are four on-shell non equivalent trivially gauge invariant
vertices having no more than four derivatives. Moreover, one their
particular combination reproduces the minimal (having no more than two
derivatives) one, constructed for the first time in \cite{Zin06}.
\item Our results are in complete agreement with those of
\cite{BDGT18} obtained in the metric-like formalism. We also compared
our results with the ones obtained recently in the stringy context
\cite{LMMS21}.
\end{itemize}
{\bf Gravitational interaction}
\begin{itemize}
\item In this case also there exist enough field redefinitions to
bring the vertex into an abelian form.
\item There exist three on-shell non-equivalent trivially gauge
invariant vertices.
\item Besides, there are two abelian vertices, which are not
equivalent to the trivially gauge invariant ones and can not be
removed by field redefinitions. Moreover, their existence is crucial
for the possibility to reproduce the minimal (with no more than two
derivatives) gravitational vertex.
\end{itemize}

Our paper is organized as follows. In Section 2 we provide all
necessary kinematical information on the frame-like gauge invariant
description of massive spin 2. Section 3 describes massive spin 2
self-interaction, while Section 4 devoted to the gravitational vertex.
A couple of appendices provides the minimal (with no more than two
derivatives) vertices obtained by direct constructive approach.

{\bf Notations and conventions}
We work in the (anti) de Sitter space with the background frame $e^a$ 
and its inverse $\hat{e}_a$. We heavily use short-hand notations for
their wedge products:
$$
E^{a[k]} = e^{a_1} \wedge e^{a_2} \wedge \dots \wedge e^{a_k},
$$
$$
\hat{E}_{a[k]} = \hat{e}_{a_1} \wedge \hat{e}_{a_2} \wedge \dots
\wedge \hat{e}_{a_k}.
$$
Here and in what follows square brackets denote antysymmetrization. A
couple of useful relations:
$$
\hat{E}_{a[k]} \wedge e^b = \delta_a{}^b \hat{E}_{a[k-1]}, 
$$
$$
\hat{E}_{a[k]} \wedge e^a = (d-k+1) \hat{E}_{a[k-1]}.
$$
An $(A)dS$ covariant derivative $D$ is defined so that
$$
D \wedge D \xi^a = - \kappa E^a{}_b \xi^b.
$$
In the main text we systematically omit the wedge product sign 
$\wedge$.

\section{Kinematics}

In this section we provide all necessary kinematical information on
the frame-like gauge invariant description of massive spin 2 in 
$(A)dS_d$ space with $d \ge 4$ \cite{Zin08b,PV10,KhZ19}.

\subsection{General formalism}

In general, to construct a gauge invariant description of massive
spin $s$ one uses a set of fields necessary for description of
massless spin $s$, $s-1$, $\dots$. So for the frame-like (first order)
gauge invariant formulation of massive spin 2 we introduce one-forms
$\Omega^{a[2]}$, $f^a$ for massless spin 2, one form $A$ and zero-form
$B^{a[2]}$ for spin 1 and zero-forms $\pi^a$, $\varphi$ for spin 0.
Then the Lagrangian, describing massive spin 2 in $(A)dS_d$ space, has
the form:
\begin{eqnarray} 
{\cal L}_0 &=& 
  \frac{1}{2} {\hat E}_{a[2]} \Omega^{ac} \Omega^{a}{}_{c}
- \frac{1}{2} {\hat E}_{a[3]} \Omega^{a[2]} D f^a + 
\frac{1}{2} B_{ab}B^{ab} \nonumber \\
 && - {\hat E}_{a[2]} B^{a[2]} D A - \frac{(d-1)(d-2)}{2} \pi_a\pi^a +
(d-1)(d-2) {\hat e}_a \pi^a D \varphi \nonumber \\
&& + m [ {\hat E}_{a[2]} \Omega^{a[2]} A + \hat{e}_a B^{ab} f_b
] - 2(d-1)M \hat e_a \pi^a A_\mu \nonumber \\
&& + \frac{M^2}{2} {\hat E}_{a[2]} f^a f^a - (d-1)mM \hat{e}_a
f^a \varphi + \frac{d(d-1)}{2} m^2 \varphi^2,
\end{eqnarray}
where
\begin{equation}
M^2 = m^2 - (d-2)\kappa.
\end{equation}
The structure of the Lagrangian is common to the gauge invariant
approach, namely, the first two lines are just the sum of the kinetic
terms for massless spin 2, 1 and 0 (to simplify the formulas we use
non-canonical normalization for spin 0); the fourth line contains all
possible mass-like terms, with the mixing terms in the third one. The
main requirement here, that completely fixes the whole  construction,
is that the Lagrangian must still be invariant under all
(appropriately modified) gauge transformations of initial massless
components. Indeed, it is not hard to check the the Lagrangian is
invariant under the following gauge transformations:
\begin{eqnarray}
\delta_0 f^a &=& D \xi^a + e_b\eta^{ba} + \frac{2m}{(d-2)}
e^a \xi, \nonumber \\
\delta_0 \Omega^{a[2]} &=& D \eta^{a[2]}
- \frac{M^2}{(d-2)} e^a \xi^{a}, \nonumber \\
\delta_0 A &=& D \xi + \frac{m}{2} \xi, \qquad
\delta_0 B^{a[2]} = - m \eta^{a[2]}, \label{gtrans_1} \\
\delta_0 \varphi &=& \frac{2M}{(d-2)} \xi, \qquad
\delta_0 \pi^a = - \frac{Mm}{(d-2)} \xi^a. \nonumber
\end{eqnarray}

One of the nice features of the frame-like formalism is that for all
fields (physical or auxiliary) one can construct gauge invariant field
strength:
\begin{eqnarray}
{\cal F}^{a[2]} &=& D \Omega^{a[2]}
- \frac{m}{(d-2)} e^a B^a - \frac{M^2}{(d-2)} e^a f^a 
+ \frac{2mM}{(d-2)}  E^{a[2]} \varphi, \nonumber \\
{\cal T}^a &=& D f^a + e_b\Omega^{ba} 
+ \frac{2m}{(d-2)} e^a A, \nonumber \\
{\cal B}^{a[2]} &=& D B^{a[2]} + m \Omega^{a[2]}
- M e^a \pi^a, \nonumber \\
{\cal A} &=& D A - \frac{1}{2}E_{a[2]}B^{a[2]}
+ \frac{m}{2}e_b f^b, \\
\Pi^a &=& D \pi^a + \frac{M}{(d-2)} e_b B^{ba}
+ \frac{Mm}{(d-2)} f^a - \frac{m^2}{(d-2)} e^a
\varphi, \nonumber \\
\Phi &=& D \varphi - e_b\pi^b - \frac{2M}{(d-2)} A.
\nonumber
\end{eqnarray}
Note that ${\cal F}^{a[2]}$, ${\cal T}^a$ and ${\cal A}$ are
two-forms, while ${\cal B}^{a[2]}$, $\Pi^a$ and $\Phi$ are one-forms.
In what follows, we will collectively call all of them as curvatures.

By straightforward calculations one can show that each curvature
satisfies a corresponding differential identity:
\begin{eqnarray}
D {\cal F}^{a[2]} &=& - \frac{m}{(d-2)} E^a{}_b {\cal B}^{ba}
+ \frac{M^2}{(d-2)} e^a {\cal T}^a
+ \frac{2mM}{(d-2)} E^{a[2]} \Phi, \nonumber \\
D {\cal T}^a &=& - e_b{\cal F}^{ba} - \frac{2m}{(d-2)}
e^a {\cal A}, \nonumber \\
D {\cal B}^{a[2]} &=& m {\cal F}^{a[2]} +
M e^a \Pi^a, \nonumber \\
D {\cal A} &=& - \frac{1}{2}E_{a[2]}{\cal B}^{a[2]}
- \frac{m}{2} e_b {\cal T}^b, \\
D \Pi^a &=& - \frac{M}{(d-2)} e_b {\cal B}^{ba}
+ \frac{Mm}{(d-2)} {\cal T}^a
+ \frac{m^2}{(d-2)} e^a \Phi, \nonumber \\
D \Phi &=& e_b \Pi^b - \frac{2M}{(d-2)} {\cal A}. \nonumber
\end{eqnarray}
In what follows "on-shell" means "on auxiliary fields equations",
i.e.:
\begin{equation}
{\cal T}^a \approx 0, \qquad {\cal A} \approx 0, \qquad
\Phi \approx0.
\end{equation}
In this case for the remaining three curvatures we obtain both
algebraic
\begin{equation}
e_b {\cal F}^{ba} \approx 0, \qquad E_{a[2]} {\cal B}^{a[2]}
\approx 0, \qquad e_b \Pi^b \approx 0, \label{alg_ident}
\end{equation}
as well as differential identities:
\begin{eqnarray}
D {\cal F}^{a[2]} &\approx& - \frac{m}{(d-2)} E^a{}_b {\cal B}^{ba},
\nonumber \\
D {\cal B}^{a[2]} &\approx& m {\cal F}^{a[2]} + M e^a \Pi^a, \\
D \Pi^a &\approx& -\frac{M}{(d-2)} e_b {\cal B}^{ba}.
\nonumber
\end{eqnarray}

Naturally, the Lagrangian equations, being gauge invariant, can be
expressed in terms of these curvatures. Indeed, a variation of the
Lagrangian under the arbitrary variations of all fields (physical and
auxiliary) has the form:
\begin{eqnarray}
\delta {\cal L}_0 &=& - \frac{1}{4} {\hat E}_{a[3]} {\cal F}^{a[2]}
\delta f^a - \frac{1}{4} {\hat E}_{a[3]} {\cal T}^a
\delta \Omega^{a[2]} \nonumber \\
 && + {\hat E}_{a[2]} {\cal B}^{a[2]} \delta A - \frac{1}{2}
{\hat E}_{a[2]} {\cal A} \delta B^{a[2]} \nonumber \\
 && - (d-1)(d-2) \hat{e}_a \Pi^a \delta \varphi
 + (d-1)(d-2) \hat{e}_a \Phi \delta \pi^a. \label{free_eq}
\end{eqnarray}

One more nice feature of the frame-like formalism is that using these
curvatures the Lagrangian can also be rewritten in the manifestly
gauge invariant  form. The most general ansatz looks like:
\begin{eqnarray}
{\cal L}_0 &=& a_1 {\hat E}_{a[4]} {\cal F}^{a[2]}
{\cal F}^{a[2]} + a_2 {\hat E}_{a[2]} {\cal B}^{ab}
{\cal B}^a{}_b + a_3 {\hat E}_{a[2]} \Pi^a \Pi^a \nonumber \\
 && + a_4 {\hat E}_{a[3]} {\cal F}^{a[2]} \Pi^a + a_5
{\hat E}_{a[3]} {\cal B}^{a[2]} {\cal T}^a + a_6
{\hat E}_{a[2]} {\cal B}^{a[2]} \Phi.
\end{eqnarray}
It appears that the general solution for the coefficients $a_{1-6}$
has two arbitrary parameters. This ambiguity is related with the two
identities (here we use the fact that the Lagrangian is determined
only up to the total derivative):
\begin{eqnarray}
0 &=& {\hat E}_{a[4]} D [ {\cal F}^{a[2]} {\cal B}^{a[2]}] \nonumber 
\\
 &=& \frac{m}{2} {\hat E}_{a[4]} {\cal F}^{a[2]}
{\cal F}^{a[2]} + \frac{8(d-3)m}{(d-2)} {\hat E}_{a[2]}
{\cal B}^{ab} {\cal B}^a{}_b + 2(d-3)M {\hat E}_{a[3]}
{\cal F}^{a[2]} \Pi^a \nonumber \\
 && + \frac{2(d-3)M^2}{(d-2)} {\hat E}_{a[3]} {\cal B}^{a[2]}
{\cal T}^a - 2(d-3)Mm {\hat E}_{a[2]} {\cal B}^{a[2]} \Phi, \\
0 &=& {\hat E}_{a[3]} D [ {\cal B}^{a[2]} \Pi^a] \nonumber \\
 &=& - \frac{2M}{(d-2)} {\hat E}_{a[2]} {\cal B}^{ab} {\cal B}^{a}{}_b
+ 2(d-2)M {\hat E}_{a[2]} \Pi^a \Pi^a + \frac{m}{2} {\hat E}_{a[3]}
{\cal F}^{a[2]} \Pi^a \nonumber \\
 && - \frac{Mm}{(d-2)} {\hat E}_{a[3]} {\cal B}^{a[2]}
{\cal T}^a + m^2 {\hat E}_{a[2]} {\cal B}^{a[2]} \Phi.
\end{eqnarray}
Thus the Lagrangian is determined up to the shifts:
\begin{equation}
{\cal L}_0 \Rightarrow {\cal L}_0 + \rho_1 {\hat E}_{a[4]} 
D [ {\cal F}^{a[2]} {\cal B}^{a[2]}] + \rho_2 {\hat E}_{a[3]} 
D [ {\cal B}^{a[2]} \Pi^a]
\end{equation}
Using this freedom we choose (the reason for such choice will become
clear later) $a_2 = a_5 = 0$. The most straightforward way to find
explicit solution for the coefficient is to consider variations of
the Lagrangian and compare them with (\ref{free_eq}). For example:
\begin{equation}
\delta_f {\cal L}_0 = \frac{Mma_4-8(d-3)M^2a_1}{(d-2)} {\hat E}_{a[3]}
{\cal F}^{a[2]} \delta f^a
+ \left[\frac{2Mma_3}{(d-2)}-4M^2a_4\right] {\hat E}_{a[2]} \Pi^a
\delta f^a,
\end{equation}
and this gives us the equations:
\begin{equation}
\frac{Mma_4-8(d-3)M^2a_1}{(d-2)} = - \frac{1}{4}, \qquad
\frac{ma_3}{(d-2)} = 2Ma_4.
\end{equation}
Taking into account variations for all other fields we find:
\begin{equation}
a_1 = - \frac{1}{32(d-3)\kappa}, \qquad
a_3 = - \frac{(d-2)}{\kappa}, \qquad
a_4 = - \frac{m}{4M\kappa}, \qquad
a_6 = - \frac{(d-2)}{2M}.
\end{equation}
Note the the coefficient $a_1$ is the same as in the purely massless
spin 2 case. 

\subsection{Partial gauge fixing}

To simplify the presentation we make a partial gauge fixing, namely we
set scalar field $\varphi$ to zero and solve its equation $\Phi
\approx 0$\footnote{We have explicitly checked that all our main
results remain to be the same.}:
\begin{equation}
\varphi = 0 \quad \Rightarrow \quad A = - \frac{(d-2)}{2M} e_a\pi^a.
\end{equation}
Note that this is possible only for $M \ne 0$ thus excluding the
so-called partially massless case, which has been considered in
\cite{Zin14}. The Lagrangian takes the form (we make rescaling 
$\pi^a \Rightarrow \frac{M}{(d-2)} \pi^a$ reflecting the fact that
$\pi^a$ plays the role of physical field now):
\begin{eqnarray}
{\cal L}_0 &=& \frac{1}{2} {\hat E}_{a[2]} \Omega^{ab} \Omega^a{}_b
- \frac{1}{2} {\hat E}_{a[3]} \Omega^{a[2]} D f^a
 + \frac{1}{2} B_{a[2]}B^{a[2]} - \hat{e}_a D B^{ab} \pi_b
\nonumber \\
 && - m \hat{e}_a \Omega^{ab} \pi_b  + m \hat{e}_a B^{ab} f_b
 + \frac{M^2}{2} {\hat E}_{a[2]} f^a f^a
 + \frac{(d-1)M^2}{2(d-2)} \pi_a\pi^a.
\end{eqnarray}
Such Lagrangian is still invariant under the remaining gauge
transformations:
\begin{eqnarray}
\delta \Omega^{a[2]} &=& D \eta^{a[2]} - \frac{M^2}{(d-2)}
e^a \xi^a, \nonumber \\
\delta f^a &=& D \xi^a + e_b\eta^{ba}, \label{gtrans_2} \\
\delta B^{a[2]} &=& - m\eta^{a[2]}, \qquad
\delta \pi^a = - m\xi^a. \nonumber
\end{eqnarray}
Moreover, for each field we still have a corresponding gauge invariant
curvature:
\begin{eqnarray}
{\cal F}^{a[2]} &=& D \Omega^{a[2]} - \frac{m}{(d-2)} E^a{}_b B^{ba} 
- \frac{M^2}{(d-2)} e^a f^a, \nonumber \\
{\cal T}^{a[2]} &=& D f^a + e_b \Omega^{ba} - \frac{m}{(d-2)} E^a{}_b
\pi^b, \nonumber \\
{\cal B}^{a[2]} &=& D B^{a[2]} + m \Omega^{a[2]}
- \frac{M^2}{(d-2)} e^a \pi^a, \\
\Pi^a &=& D \pi^a + e_b B^{ba} + m f^a. \nonumber
\end{eqnarray}
Let us stress that on-shell only ${\cal T}^a \approx 0$, while three
others are non-zero and satisfy (\ref{alg_ident}) as well as:
\begin{eqnarray}
D {\cal F}^{a[2]} &\approx& -\frac{m}{(d-2)} 
E^a{}_b {\cal B}^{ba}, \nonumber \\
D {\cal B}^{a[2]} &\approx& m {\cal F}^{a[2]}
 + \frac{M^2}{(d-2)} e^a \Pi^a, \\
D \Pi^a &\approx& -e_b {\cal B}^{ba}. \nonumber
\end{eqnarray}
As before, all variations of the Lagrangian can be expressed in terms
of curvatures:
\begin{eqnarray}
\delta {\cal L}_0 &=& - \frac{1}{4} {\hat E}_{a[3]} {\cal F}^{a[2]}
\delta f^a - \frac{1}{4} {\hat E}_{a[3]} {\cal T}^a
\delta \Omega^{a[2]} \nonumber \\
 && - \hat{e}_a {\cal B}^{ab} \delta \pi_b 
+ \hat{e}_a \Pi_b \delta B^{ab}. 
\end{eqnarray}
The most general ansatz for the Lagrangian written in terms of
curvatures now:
\begin{eqnarray}
{\cal L}_0 &=& a_1 {\hat E}_{a[4]} {\cal F}^{a[2]}
{\cal F}^{a[2]} + a_2 {\hat E}_{a[2]} {\cal B}^{ab}
{\cal B}^a{}_b + a_3 {\hat E}_{a[2]} \Pi^a \Pi^a \nonumber \\
 && + a_4 {\hat E}_{a[3]} {\cal F}^{a[2]} \Pi^a + a_5
{\hat E}_{a[3]} {\cal B}^{a[2]} {\cal T}^a. 
\end{eqnarray}
We again use the freedom to choose $a_2=a_5=0$ and obtain:
\begin{equation}
a_1 = - \frac{1}{32(d-3)\kappa}, \qquad
a_3 = \frac{16(d-3)M^2a_1}{(d-2)}, \qquad
a_4 = \frac{8(d-3)ma_1}{(d-2)}.
\end{equation}

\subsection{New variables}

From the explicit forms of the gauge transformations (\ref{gtrans_1})
or (\ref{gtrans_2}) one can see that the one-form fields play the
double role being gauge fields and (in some sense) the Stueckelberg
ones simultaneously. As a last refinement, we make a field
redefinition so that clearly separate their roles for all fields
(this, in-particular, simplifies the comparison of massless and
massive cases):
\begin{equation}
\tilde\Omega^{a[2]} = \Omega^{a[2]} - \frac{m}{(d-2)}
e^a \pi^a \quad \Rightarrow \quad
\delta \tilde\Omega^{a[2]} = D \eta^{a[2]} + \kappa e^a
\xi^a.
\end{equation}
Then the new curvatures take the form:
\begin{eqnarray}
\tilde{\cal F}^{a[2]} &=& D \tilde\Omega^{a[2]} + \kappa e^a
f^a, \nonumber \\
{\cal T}^a &=& D f^a + e_b\tilde\Omega^{ba}, \nonumber \\
{\cal B}^{a[2]} &=& D B^{a[2]} + m \tilde\Omega^{a[2]}
+ \kappa e^a \pi^a, \\
\Pi^a &=& D \pi^a + e_b B^{ba} + m f^a, \nonumber
\end{eqnarray}
so that the first two lines are exactly the same as in the massless
case. On-shell identities now:
\begin{equation}
D \tilde{\cal F}^{a[2]} \approx 0, \qquad
D {\cal B}^{a[2]} \approx m 
\tilde{\cal F}^{a[2]}
- \kappa e^a \Pi^a, \qquad
D \Pi^a \approx - e_b{\cal B}^{ba}. 
\end{equation}
At last, the Lagrangian now has a very simple form:
\begin{equation}
{\cal L}_0 = a_1 {\hat E}_{a[4]} \tilde{\cal F}^{a[2]}
\tilde{\cal F}^{a[2]} + \frac{1}{2} {\hat E}_{a[2]} \Pi^a
\Pi^a,
\end{equation}
where the first term is exactly the same as in the massless case,
while the second one is just the mass term appropriately dressed with
the Stueckelberg fields.

\section{Self-interaction} 

In this section we elaborate on the cubic vertex for the massive spin
2 self-interaction following the general pattern of the
Fradkin-Vasiliev formalism (see comparison of the massless and
massive cases in \cite{KhZ21}).

\subsection{Deformations}

The first step of the Fradkin-Vasiliev formalism is to consider the
most general consistent quadratic deformations for all gauge invariant
curvatures. In our case the most general ansatz looks like (from now
on we omit tildes on $\Omega^{a[2]}$ and ${\cal F}^{a[2]}$):
\begin{eqnarray}
\Delta {\cal F}^{a[2]} &=& b_1 \Omega^{ba}
\Omega^{a}{}_b + b_2 f^a f^a
+ b_3 \Omega^{a[2]} e_b\pi^b - b_4 e_b\Omega^{ba}
\pi^a + b_5 e^a \Omega^{ab} \pi_b \nonumber \\
 && + b_6 f^a e_b B^{ba} - b_7 e_bf^b B^{a[2]} +
b_8 e^a B^{ab} f_b + b_9 E_{b[2]} B^{ba}
B^{ba} + \frac{b_{10}}{2} B^{a[2]} E_{b[2]}B^{b[2]} \nonumber \\
 && + b_{11} E^a{}_c B^{ab} B^c{}_b + b_{12} 
E^a{}_b \pi^a  \pi^b, \nonumber \\
\Delta {\cal T}^a &=& b_{13} \Omega^{ba} 
f_b + b_{14} \Omega^{ab} e_cB^{c}{}_b - b_{15}
B^{ab} e_c\Omega^{cb} + b_{16} e^a 
\Omega^{b[2]} B_{b[2]} + b_{17} f^a e_b \pi^b \\
 && - b_{18} e_b f^b \pi^a + b_{19} e^a f^b
\pi_b + b_{20} E^{bc} B_{b}{}^a \pi_{c} + \frac{b_{21}}{2}
E_{b[2]}B^{b[2]} \pi^a + b_{22} E^a{}_b B^{bc} \pi_c, \nonumber \\
\Delta {\cal B}^{a[2]} &=& b_{23} \Omega_c{}^{a} B^{ac} +
b_{24} f^a \pi^a + b_{25} e_b B^{ba} \pi^a +
b_{26} B^{a[2]} e_b \pi^b + b_{27} e^a B^{ac} \pi_c, \nonumber \\
\Delta \Pi^a &=& b_{28} \Omega^{ab} \pi_b + b_{29} B^{ab}
f_b + b_{30} B^{ab} e^cB_{cb} + b_{31} e^a B^{b[2]} B_{b[2]}
+ b_{32} \pi^a e_b\pi^b + b_{33} e^a \pi^b \pi_b. \nonumber
\end{eqnarray}
Here consistency means that the deformed curvatures 
($\hat{\cal F}^{a[2]} = {\cal F}^{a[2]} + \Delta {\cal F}^{a[2]}$ and
so on) transform covariantly under all gauge transformations. For
example, from the explicit form of the deformation $\Delta {\cal
F}^{a[2]}$ one can directly read out the corresponding corrections to
the gauge transformations:
\begin{eqnarray}
\delta_1 \Omega^{a[2]} &=& 2b_1 \eta^{ca} \Omega^{a}{}_c -
b_3 \eta^{a[2]} e_b\pi^b- b_4 e_b \eta^{ba} \pi^{a} + b_5
e^a \eta^{ac} \pi_c \nonumber \\
 && + 2b_2 f^a \xi^a + b_6 e_b B^{ba} \xi^a
- b_7 B^{a[2]} e_b\xi^b + b_8 e^a B^{ac} \xi_c. 
\end{eqnarray}
This, in turn, means that the deformed curvature $\hat{\cal F}^{a[2]}
= {\cal F}^{a[2]} + \Delta {\cal F}^{a[2]}$ must transform as follows:
\begin{eqnarray}
\delta \hat{\cal F}^{a[2]} &=& 2b_1 \eta^{ca} 
{\cal F}^{a}{}_c + b_3 \eta^{a[2]} e_b \Pi^b
- b_4 e_b\eta^{ba} \Pi^a - b_5
e^a \eta^{ab} \Pi_b \nonumber \\
 && + 2b_2 {\cal T}^a \xi^a - b_6 
e_b {\cal B}^{ba} \xi^a - b_7 {\cal B}^{a[2]}
e_b\xi^b - b_8 e^a {\cal B}^{ab} \xi_b.
\end{eqnarray}
Such procedure nicely work in the massless case, but in the massive,
one due to the presence of one-form Stueckelberg fields, we have a lot
of possible field redefinitions to take into account:
\begin{eqnarray}
\Omega^{a[2]} &\Rightarrow& \Omega^{a[2]} + \kappa_1 B^{ba}
\Omega^a{}_b + \kappa_2 f^a \pi^a + \kappa_3 B^{a[2]}
e_b\pi^b + \kappa_4 e_b B^{ba} \pi^a + \kappa_5 e^a
B^{ac} \pi_c, \nonumber \\
f^a &\Rightarrow& f^a + \kappa_6 B^{ab} f_b +
\kappa_7 \Omega^{ab} \pi_b + \kappa_8 B^{ab} e^cB_{cb} +
\kappa_9 e^a B^{b[2]} B_{b[2]} + \kappa_{10} e_b\pi^b \pi^a +
\kappa_{11} e^a \pi^b \pi_b, \nonumber \\
\pi^a &\Rightarrow& \pi^a + \kappa_{12} B^{ab} \pi_b.  
\end{eqnarray}
To see the effect of such redefinitions, let us consider the one with
the parameter $\kappa_1$ as an illustration. It produces the following
corrections to the curvature:
\begin{eqnarray}
\frac{1}{\kappa_1} \Delta {\cal F}^{a[2]} &=& 
{\cal B}^{ca} \Omega^{a}{}_c + B^{ca} {\cal F}^{a}{}_{c} \nonumber 
\\
 && - m \Omega^{ca} \Omega^{a}{}_{c} - \kappa e_c\Omega^{ca} \pi^{a}
+ \kappa e^a \Omega^{ab} \pi_b + \kappa e_b B^{ba} f^a
 - \kappa e^a B^{ab} f_b.
\end{eqnarray}
This is equivalent to the shifts of the deformation parameters:
\begin{eqnarray}
b_1 &\Rightarrow& b_1 - m\kappa_1 \nonumber \\
b_4 &\Rightarrow& b_4 + \kappa \kappa_1 \nonumber \\
b_5 &\Rightarrow& b_5 + \kappa \kappa_1 \\
b_6 &\Rightarrow& b_6 - \kappa \kappa_1 \nonumber \\
b_8 &\Rightarrow& b_8 - \kappa \kappa_1 \nonumber
\end{eqnarray}
and introduction of the abelian deformations
\begin{equation}
\Delta {\cal F}^{a[2]} \sim {\cal B}^{ca} \Omega^a{}_c + B^{ca}
{\cal F}^a{}_c
\end{equation}
as well as the corresponding abelian corrections to the gauge
transformations:
\begin{equation}
\delta_1 \Omega^{ab} \sim {\cal B}^{ca} \eta^a{}_c
\end{equation}

By straightforward but rather lengthy calculations we have obtained
all the equations on the deformation parameters $b_{1-33}$ which
follows from the consistency conditions. Moreover, we have shown that
all these equations are invariant under the all shifts generated by
the field re-definitions $\kappa_{1-12}$ which serves as a quite 
non-trivial check for our calculations. We have found that the number
of free parameters in the general solution for the obtained equations
is equal to the number of possible field redefinitions. Moreover, by
these redefinitions all the parameters of the non-abelian
deformations can be set to zero. This leaves us with the abelian
deformations only in agreement with the general statement of
\cite{BDGT18}.

\subsection{Abelian vertices}

Recall that by abelian vertices we mean the cubic vertices which
contain two gauge invariant curvatures and one explicit field which
can be one-form $\Omega$, $f$ or zero-form $B$, $\pi$. Any cubic
vertex with one-form by making substitutions
\begin{eqnarray}
\Omega^{a[2]} &\Rightarrow& \frac{1}{m} [ {\cal B}^{a[2]} -
D B^{a[2]} - \kappa e^a \pi^a ], \nonumber \\
f^a &\Rightarrow& \frac{1}{m} [ \Pi^a - D \pi^a -
e_b B^{ba}]. 
\end{eqnarray}
integrating by parts and using differential identities for curvatures,
can be reduced to the combinations of trivially gauge invariant
vertices and some abelian ones containing zero-forms. As for the
later, any such gauge invariant vertex can be removed by field
redefinitions. Let us consider zero-form $B^{a[2]}$ as an example. 
Due to the two on-shell identities:
\begin{eqnarray}
0 &\approx& {\hat E}_{a[4]} {\cal F}^{a[2]} B^{a[2]} e_b\Pi^b
= - 2 {\hat E}_{a[3]} [ {\cal F}^{a[2]} B^{ab} \Pi_b +
{\cal F}^{ab} B^{a[2]} \Pi_b], \nonumber \\ 
0 &\approx& {\hat E}_{a[4]} e_b {\cal F}^{ba} B^{a[2]} \Pi^a
= {\hat E}_{a[3]} {\cal F}^{ad} [ B^{a[2]} \Pi_d - 2 B^a{}_b
\Pi^a],
\end{eqnarray}
there is only one independent vertex:
\begin{equation}
{\cal L}_a \sim {\hat E}_{a[3]} {\cal F}^{a[2]} B^{ab} \Pi_b. 
\end{equation}
Its variation gives:
\begin{equation}
\delta {\cal L}_a \sim m {\hat E}_{a[3]} {\cal F}^{a[2]} \eta^{ab}
\Pi_b
\end{equation}
and can be compensated by the correction to the gauge transformations
(showing its abelian nature):
\begin{equation}
\delta f^a \sim \eta^{ab} \Pi_b.
\end{equation}
At the same time, this vertex can be removed by the field redefinition
\begin{equation}
f^a \Rightarrow f^a +  B^{ab} \Pi_b.
\end{equation}
The situation with the abelian vertices with $\pi^a$ is similar. Net
result is that all abelian vertices are equivalent to the trivially
gauge invariant ones and/or can be removed by field redefinitions
again in agreement with \cite{BDGT18}.

\subsection{Trivially invariant terms}

Now let us turn to the last remaining possibility, namely, to the
trivially gauge invariant vertices. We have three curvatures 
${\cal F}$, ${\cal B}$ and $\Pi$ which do not vanish on-shell, so the
most general ansatz is:
\begin{eqnarray}
{\cal L}_1 &=& h_1 {\hat E}_{a[6]} {\cal F}^{a[2]}
{\cal F}^{a[2]} {\cal F}^{a[2]}
+ h_2 {\hat E}_{a[5]} {\cal F}^{a[2]} {\cal F}^{a[2]}
\Pi^a \nonumber \\
 && + h_3 {\hat E}_{a[4]} {\cal F}^{a[2]} \Pi^a
\Pi^a + h_4 {\hat E}_{a[3]} \Pi^a \Pi^a \Pi^a
\nonumber \\
 && + {\hat E}_{a[4]} [ h_5 {\cal F}^{a[2]} {\cal B}^{ab}
{\cal B}^a{}_b + h_6 {\cal F}^{ab}
{\cal B}_b{}^a {\cal B}^{a[2]}] \nonumber \\
 && + {\hat E}_{a[3]} [ h_7 {\cal B}^{ab} {\cal B}^a{}_b
\Pi^a + h_8 {\cal B}^{a[2]} {\cal B}^{ab}
\Pi_b].
\end{eqnarray}
But these terms are not independent. Once again using the fact that
the Lagrangian is defined only up to total derivative and also
differential identities for the curvatures we obtain:
\begin{eqnarray}
0 &=& {\hat E}_{a[6]} D [ {\cal F}^{a[2]}
{\cal F}^{a[2]} {\cal B}^{a[2]}] \nonumber \\
 &\approx& \frac{m}{2} {\hat E}_{a[6]} {\cal F}^{a[2]}
{\cal F}^{a[2]} {\cal F}^{a[2]}
 - (d-5)\kappa {\hat E}_{a[5]} {\cal F}^{a[2]}
{\cal F}^{a[2]} \Pi^a, \\
0 &=& {\hat E}_{a[5]} D [ {\cal F}^{a[2]}
{\cal B}^{a[2]} \Pi^a] \nonumber \\
 &\approx& \frac{m}{2} {\hat E}_{a[5]} {\cal F}^{a[2]}
{\cal F}^{a[2]} \Pi^a
- 2(d-4)\kappa {\hat E}_{a[4]} {\cal F}^{a[2]} \Pi^a
\Pi^a \nonumber \\
 && - 2 {\hat E}_{a[4]} [ {\cal F}^{a[2]} {\cal B}^{ab}
{\cal B}^{a}{}_b + {\cal F}^{ab}
{\cal B}_b{}^a {\cal B}^{a[2]}], \\
0 &=& {\hat E}_{a[4]} D [ {\cal B}^{a[2]} \Pi^a
\Pi^a] \nonumber \\
 &\approx& \frac{m}{2} {\hat E}_{a[4]} {\cal F}^{a[2]} \Pi^a
\Pi^a - 2(d-3)\kappa {\hat E}_{a[3]} \Pi^a \Pi^a \Pi^a \nonumber
\\
 && - 2 {\hat E}_{a[3]} [ 2 {\cal B}^{ab} {\cal B}^a{}_b
\Pi^a - {\cal B}^{a[2]} {\cal B}^{ab}
\Pi_b], \\
0 &=& {\hat E}_{a[4]} D [ {\cal B}^{a[2]} {\cal B}^{ab}
{\cal B}^{a}{}_b ] \nonumber \\
 &\approx& \frac{m}{2} {\hat E}_{a[4]} [ {\cal F}^{a[2]}
{\cal B}^{ab} {\cal B}^a{}_b + 2 
{\cal F}^{ab} {\cal B}^a{}_b {\cal B}^{a[2]}
] \nonumber \\
 && - 2\kappa {\hat E}_{a[3]} [(d-1) {\cal B}^{ab} 
{\cal B}^a{}_b \Pi^a - (d-4) {\cal B}^{a[2]}
{\cal B}^{ab} \Pi_b ].
\end{eqnarray}
Thus we have only four independent terms and we choose:
\begin{equation}
{\cal L}_1 = h_5 {\hat E}_{a[4]} {\cal F}^{a[2]}
{\cal B}^{ab} {\cal B}^a{}_b + {\hat E}_{a[3]} 
[ h_7 {\cal B}^{ab} {\cal B}^{a}{}_b \Pi^a + h_8
{\cal B}^{a[2]} {\cal B}^{ab} \Pi_b + h_4
\Pi^a \Pi^a \Pi^a].
\end{equation}
To understand the physical meaning of these results, it is convenient
to consider them in the unitary gauge $B^{a[2]} = 0$, $\pi^a=0$, where
\begin{equation}
{\cal B}^{a[2]} = m \Omega^{a[2]}, \qquad
\Pi^a = m f^a.
\end{equation}
Let us stress that all the possible field redefinitions we discussed
above necessarily contain Stueckelberg fields and so do not change
this part of the Lagrangian. We obtain:
\begin{eqnarray}
{\cal L}_1 &\approx& m^2h_5 {\hat E}_{a[4]} {\cal F}^{a[2]}
\Omega^{ab} \Omega^a{}_b \nonumber \\
 && + m^3 {\hat E}_{a[3]} [ h_7 \Omega^{ab} \Omega^{a}{}_b
f^a + h_8 \Omega^{a[2]}\Omega^{ab} f_b +
h_4 f^a f^a f^a ].
\end{eqnarray}
It is instructive to compare these expression with the minimal two
derivative vertex (see appendix A.1) which in  the unitary gauge has
the
form:
\begin{equation}
{\cal L}_1 \sim {\hat E}_{a[3]} [ \Omega^{ab} \Omega^{a}{}_b
f^a + \Omega^{a[2]} \Omega^{ab} f_b -
M^2 f^a f^a f^a].
\end{equation}

\Yautoscale0
\Yboxdim4pt

Our results are in a complete agreement with the results of
\cite{BDGT18}. Indeed, in the metric-like formalism there exists only
one main gauge invariant object (here we follow the notations of the
original paper):
$$
H_{\mu\nu} = h_{\mu\nu} + \frac{1}{m} \nabla_{(\mu} B_{\nu)} -
\frac{1}{\mu m} \nabla_\mu \nabla_\nu \varphi -
\frac{2m}{\mu(D-2)} g_{\mu\nu} \varphi
$$
(compare with one-form $\Pi^a$), but there exist also two gauge
invariant objects which do not contain more than two derivatives of
the physical field:
$$
H_{\mu\nu\rho}^{\yng(2,1)} = \nabla_\mu H_{\nu\rho} -
\nabla_\nu H_{\mu\rho}
$$
$$
H_{\mu\nu\rho\sigma}^{\yng(2,2)} = \nabla_\mu \nabla_{[\rho} 
H_{\sigma]\mu} - \nabla_\nu \nabla_{[\rho} H_{\sigma]\mu}
+ \nabla_\rho \nabla_{[\mu} h_{\nu]\sigma} - \nabla_\sigma
\nabla_{[\mu} H_{\nu]\rho}
$$
(compare with one form ${\cal B}^{a[2]}$ and two-form 
${\cal F}^{a[2]}$). The authors of \cite{BDGT18} have shown that
besides the minimal two derivative vertex there exist just three
vertices which do not contain more than four derivatives (here we
provide slightly different but equivalent form just to stress the
similarity with our frame-like results):
\begin{eqnarray*}
\bar{a}_0^{(dRGT)} &\sim& \left\{ 
\phantom{|}^{\mu_1\mu_2\mu_3}_{\nu_1\nu_2\nu_3} \right\}
H_{\mu_1}{}^{\nu_1} H_{\mu_2}{}^{\nu_2} H_{\mu_3}{}^{\nu_3}, \\
\bar{a}_0^{(PL_1)} &\sim& \left\{ 
\phantom{|}^{\mu_1\mu_2\mu_3\mu_4}_{\nu_1\nu_2\nu_3\nu_4} \right\}
H^{\yng(2,2)}{}_{\mu_1}{}^{\nu_1}{}_{\mu_2}{}^{\nu_2} 
H_{\mu_3}{}^{\nu_3} H_{\mu_4}{}^{\nu_4}, \\
\bar{a}_0^{(PL_2)} &\sim& \left\{ 
\phantom{|}^{\mu_1\mu_2\mu_3\mu_4\mu_5}_{\nu_1\nu_2\nu_3\nu_4\nu_5}
\right\} H^{\yng(2,2)}{}_{\mu_1}{}^{\nu_1}{}_{\mu_2}{}^{\nu_2}
H^{\yng(2,2)}{}_{\mu_3}{}^{\nu_3}{}_{\mu_4}{}^{\nu_4} 
H_{\mu_5}{}^{\nu_5}.
\end{eqnarray*}
The first one is related with the so-called dRGT gravity
\cite{RGT10,RGT11}, while the other two are related with the so-called
pseudo-linear terms \cite{Hin13,BHJ18}. The authors also noted that
the vertices $\bar{a}_0^{(PL_{1,2})}$ can be rewritten in terms of
$H^{\yng(2,1)}$ instead of $H^{\yng(2,2)}$, so that there exist
differential identities for these objects similar to the ones given at
the beginning of this subsection.

Recently,, an interesting paper \cite{LMMS21} appeared, where the
cubic vertices for massive spin 2 self-interaction as well  as its
interaction with massless graviton were elaborated in the stringy
context. The massive spin 2 comes from the first massive level of open
bosonic or superstring, while massless graviton comes from the closed
(super)string. The aim of the paper was to compare the stringy results
with those of the bigravity theory \cite{HR11}. To this purpose, the
authors calculated three point amplitudes and then converted them into
a so-called transverse-traceless parts of the cubic vertices. They
obtained (here we also follow the notations of the original paper;
in-particular, symmetric tensor $M_{\mu\nu}$ corresponds to the
massive spin 2):
\begin{itemize}
\item bigravity
\begin{eqnarray*}
{\cal L}_{M^3} &=& 
\frac{(-\beta_1+\beta_3)(1+\alpha^2)^{3/2}m_g}{6\alpha} [M^3] \\
 && + \frac{(1-\alpha^2)}{m_g\alpha\sqrt{1+\alpha^2}}
M^{\mu\nu} ( \partial_\mu M_{\rho\sigma} \partial_\nu M^{\rho\sigma}
- 2 \partial_\nu M_{\rho\sigma} \partial^\sigma M_\mu{}^\rho ),
\end{eqnarray*}
\item superstring
\begin{eqnarray*}
{\cal L}_{M^3}^{eff} &=& \frac{g_0}{\alpha'} \left\{ [M^3] 
+ 2\alpha' M^{\mu\nu} [ \partial_\mu M_{\rho\sigma} \partial_\nu
M^{\rho\sigma} - 3 \partial_\nu M_{\rho\sigma} \partial^\sigma
M_\mu{}^\rho] \right. \\
 && \qquad \qquad \left. + 4\alpha'^2 \partial^\mu \partial^\nu
M_{\rho\sigma} \partial^\rho M_\nu{}^\kappa \partial^\sigma
M_{\rho\kappa} \right\}, 
\end{eqnarray*}
\item bosonic string
\begin{eqnarray*}
{\cal L}_{M^3}^{eff,bos} &=& \frac{g_0}{\alpha'} \left\{ 2[M^3] 
+ 3\alpha' M^{\mu\nu} [ \partial_\mu M_{\rho\sigma} \partial_\nu
M^{\rho\sigma} - 4 \partial_\nu M_{\rho\sigma} \partial^\sigma
M_\mu{}^\rho] \right. \\
 && \left. + 2\alpha'^2 \partial^\mu \partial^\nu
M_{\rho\sigma} \partial^\rho M_\nu{}^\kappa \partial^\sigma
M_{\rho\kappa} - 2\alpha'^3 \partial^\mu \partial^\nu
M_{\mu\nu} \partial^\rho \partial^\sigma M_{\rho\sigma}
\partial^\lambda \partial^\kappa M_{\lambda\kappa} \right\}.
\end{eqnarray*}
\end{itemize}
Thus the general structure is indeed the same (one term without
derivatives, two terms with two derivatives and so on), while the
relative coefficients are different in all three cases.

\subsection{Non-abelian version}

In the kinematical section we have seen that the free Lagrangian for
massive spin 2 can be written as the massless one plus mass term
appropriately dressed with the Stueckelberg fields. At the same time,
the cubic vertex in its trivially gauge invariant form drastically
differs from that of the massless case being non-abelian in nature. In
this subsection we reconstruct our vertex in the form as close to the
massless one as possible. For this purpose, we start with the
following ansatz for the deformation of the massive spin 2 curvatures:
\begin{eqnarray}
\Delta {\cal F}^{a[2]} &=& b_1 \Omega^{ba}
\Omega^{a}{}_b + b_2 f^a f^a, \nonumber \\
\Delta {\cal T}^a &=& b_3 \Omega^{ab}
f_b, \nonumber \\
\Delta {\cal B}^{a[2]} &=& b_4 B^{ba} \Omega^a{}_b
+ b_5 f^a \pi^a, \\
\Delta \Pi^a &=& b_6 \Omega^{ab} \pi_b 
+ b_7 B^{ab} f_b \nonumber
\end{eqnarray}
and require that the deformed curvatures transform covariantly.\\\
{\bf $\eta^{ab}$-transformations} In this case corrections to the
gauge transformations have the form:
\begin{equation}
\delta \Omega^{a[2]} = - b_1 \eta^{ba} \Omega^a{}_b, \quad
\delta f^a = - b_3 \eta^{ab} f_b, \quad
\delta B^{ab} = b_4 \eta^{ba} B^a{}_b, \quad
\delta \pi^a = - b_6 \eta^{ab} \pi_b,
\end{equation}
while the variations of the deformed curvatures are:
\begin{eqnarray}
\delta \hat{\cal F}^{ab} &=& - b_1 \eta^{ba} D
\Omega^{a}{}_b + b_2 f^a e_b\eta^{ba} 
 - \kappa b_3 e^a \eta^{a}{}_b f^b, \nonumber \\
\delta \hat{\cal T}^a &=& - b_3 \eta^{ab} D
f_b + (b_3-b_1) \Omega^a{}_b e_c\eta^{cb}
- b_1 \eta^{ab} e_c\Omega^c{}_b, \\
\delta \hat{\cal B}^{a[2]} &=& b_4 \eta^{ba} D B^a{}_b 
- m(b_1+b_4) \eta^{ba} \Omega^{a}{}_b + b_5 e_b \eta^{ba}
\pi^a - \kappa b_6 e^a \eta^{ab} \pi_b, \nonumber \\
\delta \hat\Pi^a &=& - b_6 \eta^{ab} D \pi_b + (b_7-b_4)
B^{ab} e_c\eta^{c}{}_b - m(b_3+b_7) \eta^{ab} f_b
+ b_4 \eta^{ab} e_c B^c{}_b. \nonumber
\end{eqnarray}
For consistency this must coincide with
\begin{eqnarray}
- b_1 \eta^{ba} {\cal F}^{a}{}_b &=& - b_1 \eta^{ba} 
D \Omega^{a}{}_b + \kappa b_1 f^a e_b \eta^{ba} - \kappa b_1 e^a
\eta^{ab} f_b, \nonumber \\
- b_3 \eta^{ab} {\cal T}_b &=& - b_3 \eta^{ab}
[ D f_b + e_c\Omega^c{}_b ], \\
b_4 \eta^{ba} {\cal B}^a{}_b &=& b_4 \eta^{ba} D B^{a}{}_b +
mb_4 \eta^{ba} \Omega^a{}_b - \kappa b_4 \eta^a
\pi^a + \kappa b_4 e^a \eta^{ab} \pi_b, \nonumber  \\
- b_6 \eta^{ab} \Pi_b &=& - b_6 \eta^{ab} [ D \pi_b +
e^c B_{cb} + m f_b ], \nonumber
\end{eqnarray}
which gives us:
\begin{equation}
b_2 = \kappa b_1, \qquad b_3 = b_1, \qquad
b_4 = - \frac{b_1}{2}, \qquad b_5 = - \kappa b_4, \qquad
b_6 = - b_4, \qquad b_7 = b_4. 
\end{equation}
{\bf $\xi^a$-transformations} Here the corrections look like:
\begin{equation}
\delta \Omega^{a[2]} = b_2 f^a \xi^a, \quad
\delta f^a = b_3 \Omega^{ab} \xi_b, \quad
\delta B^{a[2]} =  b_5 \pi^a \xi^a, \quad
\delta \pi^a = - b_7 B^{ab} \xi_b,
\end{equation}
while the variations have the form:
\begin{eqnarray}
\delta {\cal F}^{a[2]} &=& b_2 D f^a
\xi^a + \kappa b_1 e_b\Omega^{ba} \xi^a +
\kappa (b_3-b_1) e^a \Omega^{ab} \xi_b, \nonumber \\
\delta \hat{\cal T}^a &=& b_3 D 
\Omega^{ab} \xi_b + \kappa b_3 e^a f^b
\xi_b - (\kappa b_3- b_2) e_bf^b \xi^a + b_2 f^a
e_b\xi^b, \\
\delta \hat{\cal B}^{a[2]} &=& b_5 D \pi^a \xi^a 
 + \kappa (b_4-b_7) B^{ba} e^a \xi_b - \kappa b_4
e_bB^{ba} \xi^a + m(b_2-b_5) f^a \xi^a, \nonumber \\
\delta \hat\Pi^a &=& - b_7 D B^{ab} \xi_b + m(b_3-b_6)
\Omega^{ab} \xi_b + \kappa b_6 e^a \pi^b\xi_b
+ (b_5 - \kappa b_6) e_b\pi^b \xi^a - b_5 \pi^a e_b\xi^b. \nonumber
\end{eqnarray}
Comparing them with
\begin{eqnarray}
b_2 {\cal T}^a \xi^a &=& b_2 [ D f^a +e_b \Omega^{ba} \xi^a, \nonumber
\\
b_3 {\cal R}{}^{ab} \xi_b &=& b_3 D 
\Omega^{ab} \xi_b + \kappa b_3 e^a f^b
\xi_b + \kappa b_3 f^a e_b\xi^b, \\
b_5 \Pi^a \xi^a &=& b_5 [ D \pi^a +
e_b B^{ba} + m f^a] \xi^{a}, \nonumber \\
- b_7 {\cal B}^{ab} \xi_b &=& - b_7 D B^{ab} \xi_b - mb_7
\Omega^{ab} \xi_b - \kappa b_7 e^a \pi^b \xi_b + \kappa b_7
\pi^a e_b\xi^b \nonumber
\end{eqnarray}
we obtain the same solution for the coefficients $b_{1-7}$. 
Thus we obtain (we change $b_1 \to 2b_0$):
\begin{eqnarray}
\Delta {\cal F}^{a[2]} &=& 2b_0 \Omega^{ba}
\Omega^{a}{}_b + 2\kappa b_0 f^a f^a,
\nonumber \\
\Delta {\cal T}^a &=& 2b_0 \Omega^{ab}
f_b, \nonumber \\
\Delta {\cal B}^{a[2]} &=& - b_0 B^{ba} \Omega^{a}{}_b
+ \kappa b_0 f^a \pi^a, \\
\Delta \Pi^a &=& b_0 \Omega^{ab} \pi_b 
- b_0 B^{ab} \tilde{f}_b. \nonumber
\end{eqnarray}
In this, the variations of the deformed curvatures which do not vanish
on-shell have the form:
\begin{eqnarray}
\delta \hat{\cal F}^{a[2]} &\approx& 2b_0 
{\cal F}^{ba} \eta^a{}_b, \nonumber \\
\delta \hat{\cal B}^{a[2]} &\approx& - b_0 \eta^{ba} B^a{}_b
+ \kappa b_0 \Pi^a \xi^a, \\
\delta \hat\Pi^a &\approx& - b_0 \eta^{ab} \Pi_b + b_0
{\cal B}^{ab} \xi_b. \nonumber  
\end{eqnarray}
Now as in the massless cases we take the free Lagrangian but with
initial curvatures replaced by the deformed ones and add possible
abelian terms having no more than four derivatives:
\begin{eqnarray}
{\cal L} &=& a_1 {\hat E}_{a[4]} \hat{\cal F}^{a[2]}
\hat{\cal F}^{a[2]} + d_1 {\hat E}_{a[5]}
{\cal F}^{a[2]} {\cal F}^{a[2]} f^a
\nonumber \\
 && + \frac{1}{2} {\hat E}_{a[2]} \hat\Pi^a \hat\Pi^a 
+ d_2 {\hat E}_{a[3]} \Pi^a \Pi^a f^a.
\end{eqnarray}
Note that the term with the coefficient $d_1$ (which exists in $d > 4$
only) is gauge invariant by itself and the first line is the same as
in the massless spin 2 case. As for the second line, its variation
looks like
\begin{equation}
\delta {\cal L}_1 = (2d_2-b_0) {\hat E}_{a[2]} \Pi^a \eta^{ab}
\Pi_b  + (b_0-2d_2) {\hat E}_{a[2]} {\cal B}^{ab} \Pi^a
\xi_b + 2d_2 {\hat E}_{a[2]} {\cal B}^{ab} \Pi_b \xi^a.
\end{equation}
Thus we have to put $d_2 = \frac{b_0}{2}$, while with the help of
identity
$$
0 \approx {\hat E}_{a[3]} {\cal B}^{a[2]} e_b\Pi^b \xi^a = 
- {\hat E}_{a[2]} [{\cal B}^{a[2]} \Pi^c \xi_c - 2 {\cal B}^{ab}
\Pi_b \xi^a] 
$$
we find that the last term vanish on-shell.

\section{Gravitational interaction}

In this section we consider gravitational interaction of massive spin
2 in the same framework. At first, we provide the kinematics of
massless spin 2 in our notations.

\subsection{Graviton}

We describe massless graviton with the one-forms $h^a$ and 
$\omega^{a[2]}$ and the free Lagrangian:
\begin{equation}
{\cal L}_0 = \frac{1}{2} {\hat E}_{a[2]}\omega^{ab} \omega^a{}_b
- \frac{1}{2} {\hat E}_{a[3]} \omega^{a[2]} D h^a
- \frac{(d-2)\kappa}{2} {\hat E}_{a[2]} h^a h^a.
\end{equation}
This Lagrangian is invariant under the following gauge
transformations:
\begin{equation}
\delta_0 \omega^{a[2]} = D \hat\eta^{a[2]} + \kappa e^a
\hat\xi^a, \qquad
\delta_0 h^a = D \hat\xi^a + e_b\hat\eta^{ba}.
\end{equation}
Two main gauge invariant objects (curvature and torsion) are:
\begin{eqnarray}
R^{a[2]} &=& D \omega^{a[2]} + \kappa e^a h^a, \nonumber \\
T^a &=& D h^a + e_b\omega^{ba}.
\end{eqnarray}
They satisfy the usual differential identities:
\begin{equation}
D R^{a[2]} = - \kappa e^a T^a, \qquad
D T^a = - e_bR^{ba},
\end{equation}
while on-shell we have:
\begin{equation}
T^a \approx 0 \quad \Rightarrow \quad 
e_b R^{ba} \approx 0, \quad D R^{a[2]} \approx 0.
\end{equation}
At last, the free Lagrangian can be rewritten as
\begin{equation}
{\cal L}_0 = a_0 {\hat E}_{a[4]} R^{a[2]} R^{a[2]},
\qquad a_0 = - \frac{1}{32(d-3)\kappa}.
\end{equation}

\subsection{Deformations}

We begin with the deformations for the massless spin 2. The most
general ansatz has the form:
\begin{eqnarray}
\Delta R^{a[2]} &=& c_1 \Omega^{ba} \Omega^a{}_b + c_2 f^a f^a
+ c_3 \Omega^{a[2]} e_b\pi^b - c_4 e_b\Omega^{ba}
\pi^{a} + c_5 e^a \Omega^{ab} \pi_b \nonumber \\
 && + c_6 f^a e_b B^{ba} - c_7 e_bf^b B^{a[2]} +
c_8 e^a B^{ab} f_b + c_9 E_{b[2]} B^{ba} B^{ba} + \frac{c_{10}}{2}
B^{a[2]} E_{b[2]} B^{b[2]} \nonumber \\
 && + c_{11} E^{ab} B^{ac} B_{bc} + c_{12} 
E^{ab} \pi^a \pi_b, \\
\Delta T^a &=& c_{13} \Omega^{ab} 
f_b + c_{14} \Omega^{ab} e_c B^{cb} - c_{15}
B^{ab} e_c\Omega^{cb} + c_{16} e^a 
\Omega^{b[2]} B_{a[2]} + c_{17} f^a e_b \pi^b \nonumber \\
 && - c_{18} e_b f^b \pi^a + c_{19} e^a f^b
\pi_b + c_{20} E_{bc} B^{ba} \pi^c + \frac{c_{21}}{2} E_{b[2]}B^{b[2]}
\pi^a + c_{22} E^{ab} B_{bc} \pi^c. \nonumber
\end{eqnarray}
Here we also have a lot of possible field redefinitions:
\begin{eqnarray}
\omega^{a[2]} &\Rightarrow& \omega^{a[2]} + \kappa_1 B^{ba}
\Omega^a{}_b + \kappa_2f^a \pi^a + \kappa_3 B^{a[2]}
e_b\pi^b + \kappa_4 e_b B^{ba} \pi^{a} + \kappa_5 e^a
B^{ab} \pi_b, \\
h^a &\Rightarrow& h^a + \kappa_6 B^{ab}f_b +
\kappa_7 \Omega^{ab} \pi_b + \kappa_8 B^{ab} e^cB_{cb} +
\kappa_9 e^a B^{b[2]} B_{b[2]} + \kappa_{10} e_b\pi^b \pi^a +
\kappa_{11} e^a \pi^b \pi_b. \nonumber
\end{eqnarray}
The most general ansatz for the massive curvatures deformations is:
\begin{eqnarray}
\Delta {\cal F}^{a[2]} &=& b_1 \Omega^{ba}
\omega^a{}_b + b_2 h^a f^a
+ b_3 \omega^{a[2]} e_b\pi^b - b_4 e_b\omega^{ba}
\pi^{a} + b_5 e^a \omega^{ac} \pi_c \nonumber \\
 && + b_6 h^a e_bB^{ba} - b_7e_b h^b B^{a[2]} +
b_8 e^a B^{ab} h_b, \nonumber \\
\Delta {\cal T}^a &=& b_9 \Omega^{ab} h_b
+ b_{10} \omega^{ab} f_b + b_{11} 
\omega^{ab} e^c B_{cb} - b_{12} B^{ab} 
e_c \omega^{cb}+ b_{13} e^a \omega^{b[2]}
B_{b[2]} \nonumber \\
 && + b_{14} h^a e_b\pi^b - b_{15} e_b h^b \pi^a
+ b_{16} e^a h^b \pi_b, \\
\Delta {\cal B}^{a[2]} &=& b_{17} \omega^{ba} B^a{}_b +
b_{18} h^a \pi^a, \nonumber \\
\Delta \Pi^a &=& b_{19} \omega^{ab} \pi_b + b_{20}
B^{ab} h_b, \nonumber
\end{eqnarray}
with the possible fields redefinitions:
\begin{eqnarray}
\Omega^{a[1]} &\Rightarrow& \Omega^{a[1]} + \rho_1 B^{ba}
\omega^a{}_b + \rho_2 h^a \pi^a, \nonumber \\
f^a &\Rightarrow&f^a + \rho_3 B^{ab} h_b + \rho_4
\omega^{ab} \pi_b.
\end{eqnarray}
In both cases we have obtained all the equations on the deformation
parameters which follow from the consistency requirement and checked
that all of them are invariant under the shifts generated by all field
redefinitions. In both cases the number of the free parameters in the
general solutions are the same as the number of the fields
redefinitions so that all the non-abelian deformations can be set to
zero.

\subsection{Trivially invariant terms}

In this case it is convenient to proceed with the trivially invariant
terms and then consider the abelian ones. The mos general ansatz:
\begin{eqnarray}
{\cal L} &=& h_1 {\hat E}_{a[6]} R^{a[2]}
{\cal F}^{a[2]} {\cal F}^{a[2]} + h_2
{\hat E}_{a[5]} R^{a[2]} {\cal F}^{a[2]} \Pi^a
+ h_3 {\hat E}_{a[4]} R^{a[2]} \Pi^a \Pi^a \nonumber \\
 && + {\hat E}_{a[4]} [ h_4 R^{a[2]} {\cal B}^{ab}
{\cal B}^a{}_b + h_5 R^{ab} {\cal B}^{a}{}_b
{\cal B}^{a[2]} ].
\end{eqnarray}
Here we also have a couple of identities:
\begin{eqnarray}
0 &=& {\hat E}_{a[6]} D [ R^{a[2]}
{\cal F}^{a[2]} {\cal B}^{a[2]} ] \nonumber \\
 &\approx& \frac{m}{2} {\hat E}_{a[6]} R^{a[2]} 
{\cal F}^{a[2]} {\cal F}^{a[2]}
- 2(d-5)\kappa {\hat E}_{a[5]} R^{a[2]}
{\cal F}^{a[2]} \Pi^a, \\
0 &=& {\hat E}_{a[5]} D [ R^{a[2]} {\cal B}^{a[2]}
\Pi^a] \nonumber \\
 &\approx& \frac{m}{2} {\hat E}_{a[5]} R^{a[2]} 
{\cal F}^{a[2]} \Pi^a + 2(d-4)\kappa {\hat E}_{a[4]}
R^{a[2]} \Pi^a \Pi^a \nonumber \\
 && - 2 {\hat E}_{a[4]} [ R^{a[2]} {\cal B}^{ab}
{\cal B}^a{}_b - R^{a}{}_b {\cal B}^{ab}
{\cal B}^{a[2]} ],
\end{eqnarray}
so we choose:
\begin{equation}
{\cal L}_t = {\hat E}_{a[4]} [ h_4 R^{a[2]} {\cal B}^{ab}
{\cal B}^a{}_b + h_5 R^{ab} {\cal B}^{a}{}_b
{\cal B}^{a[2]} + h_3 R^{a[2]} \Pi^a
\Pi^a ].
\end{equation}
Note that in $d=4$ we have one additional identity:
\begin{equation}
0 = -{\hat E}_{a[5]} R^{a[2]} e_b{\cal B}^{ba}
{\cal B}^{a[2]} = 2 {\hat E}_{a[4]} [ R^{a[2]}
{\cal B}^{ab} {\cal B}^a{}_b - R^{ab}
{\cal B}^a{}_b {\cal B}^{a[2]} ].
\end{equation}
Substituting the explicit expression for $R^{a[2]}$, integrating by
parts and using differential identities for massive spin 2 curvatures,
one can show that our trivially invariant terms are equivalent to some
combinations of the abelian terms:
\begin{eqnarray}
V_3 &=& 2 {\hat E}_{a[4]} \Pi^a \Pi^a [ D
\omega^{a[2]} + 2\kappa e^a h^a] \nonumber \\
 &\approx& 4 {\hat E}_{a[3]} {\cal B}^{a}{}_b [ 2 g^{ab}\Pi^a
\omega^{ab} + \Pi^b \omega^{a[2]}]
+ 4(d-3)\kappa {\hat E}_{a[3]} \Pi^a \Pi^a h_\alpha^a, \\
V_4 &=& 2 {\hat E}_{a[4]} {\cal B}^{ab} {\cal B}^a{}_b
[ D \omega^{a[2]} + 2 \kappa e^a h^a] \nonumber \\
 &\approx&  - 2m {\hat E}_{a[4]} {\cal F}^{ab} {\cal B}^a{}_b
\omega^{a[2]} + 4\kappa {\hat E}_{a[3]} {\cal B}^{a}{}_{b}
[ 2 \Pi^a \omega^{ab} - (d-4) \Pi^b \omega^{a[2]}] \nonumber \\
 && + 4(d-3)\kappa {\hat E}_{a[3]} {\cal B}^{ab} {\cal B}^a{}_b
h^a, \\
V_5 &=& 2 {\hat E}_{a[4]} {\cal B}^{a}{}_b {\cal B}^{a[2]}
[ D \omega^{ab} + \kappa (e^a h^b - e^b h^a)] \nonumber \\
 &\approx& - m {\hat E}_{a[4]} [ {\cal F}^{ab} {\cal B}^{a[2]}
- {\cal F}^{a[2]} {\cal B}_\alpha{}^{ab}] \omega^a{}_b \nonumber \\
 && - 2\kappa [ {\hat E}_{a[3]} ( - (d-4) \Pi_b {\cal B}^{a[2]}
\omega^{ab} + 2(d-2) {\cal B}^{a}{}_b \Pi^a \omega^{ab} ) ] \nonumber
\\
 && - 2\kappa {\hat E}_{a[3]} [ (d-4) {\cal B}^{ab} 
{\cal B}^{a[2]} h_b + 2 {\cal B}^{ab}
{\cal B}^{a}{}_b h^a]. 
\end{eqnarray}

\subsection{Abelian vertices}

There are two possible types of abelian vertices: those with the
massive spin 2 components $\Omega,f,B,\pi$ and with the massless spin
2 $\omega,h$ ones. Exactly as in the massive spin 2 self-interaction
case all gauge invariant abelian vertices of the first type are
equivalent to some combinations of the trivially gauge invariant
vertices and/or can be removed by field redefinitions. So we consider
here only second type.

Taking into account on-shell identities:
\begin{eqnarray}
0 &\approx& {\hat E}_{a[5]} e_b{\cal F}^{ba} {\cal B}^{a[2]}
\omega^{a[2]} = - 2 {\hat E}_{a[4]} {\cal F}^{a}{}_b
[ {\cal B}^{ab} \omega^{a[2]} + 
{\cal B}^{a[2]} \omega^{ab} ], \nonumber \\
0 &\approx& - {\hat E}_{a[4]} {\cal B}^{a[2]} e_c \Pi^c
\omega^{a[2]} = - 2 {\hat E}_{a[3]} [{\cal B}^{a[2]}
\Pi_b \omega^{ab} + {\cal B}^{ab}
\Pi_b \omega^{a[2]} ],
\end{eqnarray}
the most general ansatz for such abelian vertices has the form (for
terms with five and four derivatives respectively);
\begin{eqnarray}
{\cal L}_{a5} &=& {\hat E}_{a[4]} [ g_1 {\cal F}^{a[2]}
{\cal B}^{ab} + g_2 {\cal F}^{ab}
{\cal B}^{a[2]} ] \omega^{a}{}_b \nonumber \\
 && + {\hat E}_{a[3]} [ g_3 {\cal B}^{a[2]} \Pi^b + g_4
{\cal B}^{ab} \Pi^a ] \omega^a{}_b, \\
{\cal L}_{a4} &=& d_1 {\hat E}_{a[5]} {\cal F}^{a[2]}
{\cal F}^{a[2]} h^a + d_2 {\hat E}_{a[4]} 
{\cal F}^{a[2]} \Pi^a h^a \nonumber \\
 && + {\hat E}_{a[3]} [ d_3 \Pi^a \Pi^a h^a + d_4
{\cal B}^{a}{}_b {\cal B}^{ab} h^a + d_5
{\cal B}^{a[2]} {\cal B}^{a}{}_b h_b ].
\end{eqnarray}
Not all of them are completely independent as can be seen from
\begin{eqnarray}
0 &=& {\hat E}_{a[5]} D [ {\cal F}^{a[2]}
{\cal B}^{a[2]} h^a] \nonumber \\
 &\approx& \frac{m}{2} {\hat E}_{a[5]} {\cal F}^{a[2]}
{\cal F}^{a[2]} h^a - 2(d-4)\kappa
{\hat E}_{a[4]} {\cal F}^{a[2]} \Pi^a h^a \nonumber \\
 && - 2 {\hat E}_{a[4]} [ {\cal F}^{a[2]} {\cal B}^{a}{}_b +
{\cal F}^{a}{}_b {\cal B}^{a[2]}] \omega^{ab}, \\
0 &=& {\hat E}_{a[4]} D [ {\cal B}^{a[2]} \Pi^a
h^a ] \nonumber \\
 &\approx& \frac{m}{2} {\hat E}_{a[4]} {\cal F}^{a[2]} \Pi^a
h^a - 2(d-3)\kappa {\hat E}_{a[3]} \Pi^a \Pi^a
h^a \nonumber \\
 && - {\hat E}_{a[3]} [ {\cal B}^{a[2]} {\cal B}^{ab}
h_b + 2 {\cal B}^{ab} {\cal B}^a{}_b
h_\alpha{}^a] \nonumber \\
 && + {\hat E}_{a[3]} [ {\cal B}^{a[2]} \Pi^b - 2
{\cal B}^{ab} \Pi^a] \omega^a{}_b.
\end{eqnarray}
This leads us the the following form of the abelian vertex which is
not equivalent to any combination of the trivially gauge invariant
ones (taking into account relations at the end of previous
subsection):
\begin{equation}
{\cal L}_a = {\hat E}_{a[3]} [ g_3 {\cal B}^{a[2]} \Pi_b
\omega^{ab} + d_3 \Pi^a \Pi^a h^a + d_4
{\cal B}^{ab} {\cal B}^a{}_b h^a + d_5
{\cal B}^{a[2]} {\cal B}^{ab} h_b].
\end{equation}
Now let us require this vertex to be gauge invariant.\\
{\bf $\hat\eta^{ab}$-transformations} We obtain:
\begin{equation}
\delta {\cal L}_a = \frac{g_3-d_4+2d_5}{2} {\hat E}_{a[2]} {\cal
B}^{a[2]} {\cal B}^{b[2]} \hat\eta_{b[2]}, 
\end{equation}
but this can be compensated by the appropriate correction to the gauge
transformation:
\begin{equation}
\delta A \sim {\cal B}_{b[2]} \hat\eta^{b[2]}.
\end{equation}
{\bf $\hat\xi^a$-transformations} These variations give us:
\begin{eqnarray}
\delta_\xi {\cal L}_a &=& - \frac{md_5}{2} {\hat E}_{a[3]} 
{\cal F}^{a[2]} {\cal B}^{ab} \hat\xi_b
 + [ (d-2)\kappa g_3 + d_3 + (d-3)\kappa(d_5-d_4)]
{\hat E}_{a[2]} {\cal B}^{a[2]} \Pi^b \hat\xi_b \nonumber \\
 && + m(d_5-d_4) {\hat E}_{a[3]} {\cal F}^{ab} 
{\cal B}^a{}_b \hat\xi^a - 2[ d_3 + \kappa d_4 -
\kappa(d-1)d_5] {\hat E}_{a[2]} {\cal B}^{ab} \Pi^a \hat\xi_b.
\end{eqnarray}
Here all terms in the first line can be compensated by the corrections
to the gauge transformations:
\begin{equation}
\delta f^a \sim {\cal B}^{ab} \hat\xi_b, \qquad
\delta A_\mu \sim \Pi^a \hat\xi_a,
\end{equation}
while for the second line to vanish we must put
\begin{equation}
d_5 = d_4, \qquad d_3 = - M^2 d_4.
\end{equation}

Thus we have three trivially gauge invariant vertices $V_{3,4,5}$ and
two independent abelian ones (with the coefficients $g_3$ and $d_4$).
The vertices $V_{4,5}$ contain higher derivative terms, while the
remaining three in the unitary gauge produce:
\begin{eqnarray}
\frac{1}{m^2} {\cal L}_1 &=& h_3 {\hat E}_{a[4]} R^{a[2]} 
f^a f^a - g_3 {\hat E}_{a[3]} \Omega^{a[2]}
\omega^{ab} f_b \nonumber \\
 && + d_4 {\hat E}_{a[3]} [ \Omega^{ab} \omega^a{}_b h^a
+ \Omega^{a[2]} \omega^{ab} h_b 
- M^2f^a f^a h^a]. 
\end{eqnarray}
Again it is instructive to compare these results with the minimal two
derivative vertex also in the unitary gauge (see appendix A.2):
\begin{eqnarray}
{\cal L}_1 &\sim& {\hat E}_{a[3]} [ \Omega^{ab} \omega^a{}_b
h^a + \Omega^{a[2]} \omega^{ab} h_b
+ \Omega^{a[2]} \omega^{ab} f_b ] \nonumber \\
 && + \frac{1}{4} {\hat E}_{a[4]} R^{a[2]} f^a f^a
- M^2 {\hat E}_{a[3]}f^a f^a h^a.
\end{eqnarray}
Note that for this vertex the two derivative part for bigravity
\cite{HR11} and from  (super)string \cite{LMMS21} coincides (here
symmetric tensor $G_{\mu\nu}$ corresponds to the massless graviton):
\begin{eqnarray*}
{\cal L}_{GM^2} &=& \frac{1}{m_g\sqrt{1+\alpha^2}} [ G^{\mu\nu}
(\partial_\mu M_{\rho\sigma} \partial_\nu M^{\rho\sigma} -
4 \partial_\nu M_{\rho\sigma} \partial^\sigma M_\mu{}^\rho) \\
 && \qquad \qquad + 2 M^{\mu\nu} (\partial_\mu G_{\rho\sigma}
\partial_\nu M^{\rho\sigma} - \partial_\rho G_{\mu\sigma}
\partial_\nu M^{\rho\sigma}) ].
\end{eqnarray*}

\subsection{Comeback}

In this subsection we reconstruct the same vertex in the non-abelian
form so that the massive theory can be considered as a deformation of
the massless one.

{\bf Massive spin 2 deformations} We consider the following restricted
ansatz
\begin{eqnarray}
\Delta {\cal F}^{a[2]} &=& b_1 \Omega^{ba}
\Omega^a{}_b + b_2\kappa f^a h^a, \nonumber \\
\Delta {\cal T}^a &=& b_3 \Omega^{ab} f_b
+ b_4 \Omega^{ab} h_b, \nonumber \\
\Delta {\cal B}^{a[2]} &=& b_5 \Omega^{ba} B^a{}_b + 
b_6\kappa h^a \pi^a, \\
\Delta \Pi^a &=& b_7 \Omega^{ab} \pi_b + b_8 B^{ab}
h_b \nonumber
\end{eqnarray}
and require it to be consistent. \\
{\bf $\hat\eta^{ab}$-transformations} Here the corrections to the
gauge transformations look like:
\begin{equation}
\delta \Omega^{a[2]} = - b_1 \hat\eta^{ba} \Omega^a{}_b,
\quad \delta f^a = - b_3 \hat\eta^{ab}f_b, \quad
\delta B^{a[2]} = - b_5 \hat\eta^{ba} B^a{}_b, \quad
\delta \pi^a = - b_7 \hat\eta^{ab} \pi_b,
\end{equation}
while the variations for the deformed curvatures are:
\begin{eqnarray}
\delta \hat{\cal F}^{a[2]} &=& - b_1 \hat\eta^{ba}
D \Omega^a{}_b + b_2\kappa f^a
e_b \hat\eta^{ba} - b_3\kappa e^a \hat\eta^{ab}
f_b, \nonumber \\
\delta \hat{\cal T}^a &=& - b_3 \hat\eta^{ab} D
f_b + (b_4-b_1) \Omega^{ab} e^c \hat\eta_{cb} 
- b_1 \hat\eta^{ab} e^c\Omega_{cb}, \nonumber \\
\delta \hat{\cal B}^{a[2]} &=& - b_5 \hat\eta^{ba} D B^{a}{}_{b}
 + b_6\kappa e_b \hat\eta^{ba} \pi^{a} - mb_1 \hat\eta^{ba}
\Omega^{a}{}_b - b_7\kappa e^a \hat\eta^{ab} \pi_b, \\
\delta \hat\Pi^a &=& - b_7 \hat\eta^{ab} D \pi_b + (b_8+b_5)
B^{ab} e^c \hat\eta_{cb}- b_5 \hat\eta^{ab} e^cB_{cb}- mb_3
\hat\eta^{ab}f_b. \nonumber 
\end{eqnarray}
Comparing them with
\begin{eqnarray}
- b_1 \hat\eta^{ba} {\cal F}^{a}_{b} &=& - b_1
\hat\eta^{ba} [ D \Omega^{a}_{b} + \kappa 
e^a f_b - \kappa e_b f^a ] \nonumber \\
 &=& - b_1 \hat\eta^{ba} [ D \Omega^{a}{}_b 
 + b_1\kappa f^a e_b\hat\eta^{ba}
 - b_1\kappa e^a \hat\eta^{ab} f_b ], \nonumber \\
- b_3 \hat\eta^{ab} {\cal T}_b &=& - b_3 \hat\eta^{a}{}_b
[ D f^b + e_c\Omega^{cb} ], \\
- b_5 \hat\eta^{ba} {\cal B}^{a}{}_b &=& - b_5 \hat\eta^{ba}
[ D B^{a}{}_b + m \Omega^{a}{}_b + \kappa e^a \pi_b
- \kappa e_b \pi^a ] \nonumber \\
 &=& - b_5 \hat\eta^{ba} D B^{a}{}_b - mb_5 \hat\eta^{ba}
\Omega^{a}{}_b + b_5\kappa e_b\hat\eta^{ba} \pi^{a}
- b_5\kappa e^a \hat\eta^{ab} \pi_b ], \nonumber \\
- b_7 \hat\eta^{ab} \Pi_b &=& - b_7 \hat\eta^{ab}
[ D \pi_b + e^cB_{cb}+ mf_b ], \nonumber 
\end{eqnarray}
we obtain:
\begin{equation}
b_2 = b_3 = b_4 = b_5 = b_6 = b_7 = - b_8 = b_1.
\end{equation}
All calculations for $\hat\xi^a$, $\eta^{ab}$  and $\xi^a$
transformations are similar and produce the same solution.
Complete variations of the deformed curvatures which are non-zero 
on-shell have the form:
\begin{eqnarray}
\delta {\cal F}^{a[2]} &\approx& - b_1 \hat\eta^{ba}
{\cal F}^{a}{}_b - b_1 \eta^{ba} R^{a}{}_b, \nonumber \\
\delta \hat{\cal B}^{a[2]} &\approx& - b_1 \hat\eta^{ba}
{\cal B}^{a}{}_b + b_1\kappa \Pi^{a} \hat\xi^{a}, \\
\delta \hat\Pi^a &\approx& - b_1 \hat\eta^{ab} \Pi_b 
 + b_1 {\cal B}^{ab} \hat\xi_b. \nonumber
\end{eqnarray}
{\bf Deformations for the graviton} are chosen to be:
\begin{equation}
\Delta R^{a[2]} = c_1 \Omega^{ba} 
\Omega^{a}{}_b + c_2 \kappa f^a f^a, \qquad
\Delta T^a = c_3 \Omega^{ab} f_b.
\end{equation}
{\bf $\eta^{ab}$-transformations} Here the corrections to the gauge
transformations are:
\begin{equation}
\delta \Omega^{ab} = - c_1 \eta^{ba} \Omega^{a}{}_b, \qquad
\delta h^a = - c_3 \eta^{ab}f_b,
\end{equation}
while the variations for the deformed curvatures have the form:
\begin{eqnarray}
\delta \hat{R}^{a[2]} &=& - c_1 \eta^{ba} D
\Omega^{a}{}_b + c_2\kappa f^a e_b \eta^{ba}
 - c_3\kappa e^a \eta^{ab} f_b, \nonumber \\
\delta \hat{T}^a &=& - c_3 \eta^{ab} D f_b
 + (c_3-c_1) \Omega^{ab} e^c \eta_{cb}- c_1 \eta^{a}{}_b
e_c\Omega^{cb}.
\end{eqnarray}
Comparing them with:
\begin{eqnarray}
- c_1 \eta^{ba} {\cal F}^{a}{}_b &=& - c_1 [ \eta^{ba}
D \Omega^{a}{}_b - \kappa f^{a} 
e_c \eta^{ca} + \kappa e^a \eta^{ab} f_b ], \nonumber \\
- c_3 \eta^{ab} {\cal T}{}_b &=& - c_3 \eta^{ab} [
D f_b - e_c \Omega^{cb} ], 
\end{eqnarray}
we obtain:
\begin{equation}
c_2 = c_3 = c_1
\end{equation}
Similarly for $\xi^a$-transformations. The only non-zero on-shell
variation is: 
\begin{equation}
\delta \hat{R}^{a[2]} \approx - c_1 \eta^{ba} 
{\cal F}^{a}{}_b.
\end{equation}
Now we write an interacting Lagrangian containing no more than four
derivatives (note that we still need some abelian terms here):
\begin{eqnarray}
{\cal L} &=& a_0 {\hat E}_{a[4]} [ \hat{R}^{a[2]} 
\hat{R}^{a[2]} + \hat{\cal F}^{a[2]}
\hat{\cal F}^{a[2]} ] + d_1 {\hat E}_{a[5]} 
{\cal F}^{a[2]} {\cal F}^{a[2]} h^a
\nonumber \\
 && + \frac{1}{2} {\hat E}_{a[2]} \hat\Pi^a \hat\Pi^a
+ d_2 {\hat E}_{a[3]} \Pi^a \Pi^a h^a.
\end{eqnarray}
Note that the first line is exactly the same as in the massless case
\cite{Zin14} and is invariant by itself provided $b_1 = c_1$, while
for the variation of the second line we obtain:
\begin{eqnarray}
\delta {\cal L} &=& b_1 {\hat E}_{a[2]} {\cal B}^{ab} \Pi^a
\hat\xi_b + 2d_2 {\hat E}_{a[3]} e_b {\cal B}^{ba} \Pi^a
\hat\xi^a \nonumber \\
 &=& (b_1-2d_2) {\hat E}_{a[2]} {\cal B}^{ab} \Pi^a \hat\xi_b 
 - 2d_2 {\hat E}_{a[2]} {\cal B}^{ab} \Pi_b \hat\xi^a
\end{eqnarray}
Thus we have to put $d_2 = \frac{b_1}{2}$, while the last term vanish
on-shell due to identity:
$$
0 \approx - {\hat E}_{a[3]} {\cal B}^{a[2]} e_b \Pi^b
\hat\xi^a = {\hat E}_{a[2]} [ {\cal B}^{a[2]} \Pi^b \hat\xi_b
- 2 {\cal B}^{ab} \Pi_b \hat\xi^a ]
$$

\section{Conclusion}

In this work we applied the Fradkin-Vasiliev formalism based on the
frame-like gauge invariant description of the massive and massless
spin 2 to the construction of the cubic interactions vertices for
massive spin 2 self-interaction as well as its gravitational
interaction. In the first case we have shown that in agreement with
the general results of \cite{BDGT18} the vertex can be reduced to the
set of the trivially gauge invariant terms. There are four such terms
which are not equivalent om-shell and do not contain more than four
derivatives. Moreover, one their particular combination reproduces the
minimal (with no more than two derivatives) vertex \cite{Zin06}. As
for the gravitational vertex, we have shown that due to the presence
of the massless spin 2  there exist two abelian vertices (besides the
three trivially gauge invariant ones) which are not equivalent to any
trivially gauge invariant terms and can not be removed by field
re-definitions. Moreover, their existence appeared to be crucial for
the possibility to reproduce the minimal two derivatives vertex.

\section*{Acknowledgements}
M.Kh. is grateful to Foundation for the Advancement of Theoretical
Physics and Mathematics "BASIS" for their support of the work.

\appendix

\section{Minimal vertices in the constructive approach}

The spin 2 is the highest spin where all the components of the
frame-like formalism enter the free Lagrangian so one can use a very
well known constructive approach. Here we provide the results of such
approach for the minimal (i.e. with no more than two derivatives)
vertices both for the self-interaction as well as for the interaction 
with the graviton. We use the so-called modified 1 and 1/2 order
formalism (see the detailed discussion and examples of the explicit
calculations in \cite{Zin14}). In short, it means that we consider
only terms which are not equivalent on-shell and also use on-shell
conditions calculating all gauge variations. As a result, we obtain
corrections to the gauge transformations for the physical fields only.

\subsection{Self-interaction}

For the massive spin 2 self-interaction we obtained the following
minimal cubic vertex:
\begin{eqnarray}
{\cal L}_1 &=& a_0 \hat{E}_{a[3]} [ \Omega^{ab} \Omega^a{}_b f^a +
\Omega^{a[2]} \Omega^{ab} f_b ] + a_1 \hat{e}_a [ B^{b[2]} B_{b[2]}
f^a + 4 B^{ab} B_{bc} f^c ] \nonumber \\
 && + a_2 \hat{e}_a [ \pi^b \pi_b f^a - 2 \pi^a \pi^b f_b ] + a_3
\varphi B_{a[2]} B^{a[2]} \nonumber \\
 && + b_1 \hat{E}_{a[3]} \Omega^{a[2]} A f^a + b_2 \hat{e}_a B^{ab}
f_b \varphi + b_3 \hat{e}_a \varphi A \pi^a \nonumber \\
 && + c_1 \hat{E}_{a[3]} f^a f^a f^a + c_2 \hat{E}_{a[2]} f^a f^a
\varphi + c_3 \hat{e}_a f^a \varphi^2 + c_4 \varphi^3,
\end{eqnarray}
where
$$
a_1 = \frac{a_0}{2}, \qquad 
a_2 = - \frac{(d-1)(d-2)a_0}{2} + \frac{(d-2)(d-4)m^2a_0}{4M^2},
\qquad a_3 = - \frac{(d-4)ma_0}{2M},
$$
$$
b_1 = \frac{(d-4)ma_0}{(d-2)}, \qquad
b_2 = 2ma_3, \qquad
b_3 = 2ma_2,
$$
$$
6c_1 = - \frac{2(2d-5)M^2-(d-4)m^2}{(d-2)}a_0 , \qquad
c_2 = (d-1)Mma_0 - \frac{(d-4)m^3a_0}{2M},
$$
$$
c_3 = - \frac{(d-1)(d+6)m^2a_0}{4} + \frac{(3d-2)(d-4)m^4a_0}{8M^2}.
$$
In-particular, these formulas show that the partially massless limit
$M \to 0$ is possible only in $d=4$. Corrections to the physical
fields gauge transformations have the form:
\begin{eqnarray}
\delta_1 f^a &=& 2a_0 \eta^{ab} f_b - 2a_0 \Omega^{ab} \xi_b + 2b_1
f^a \xi - 2b_1 A \xi^a, \nonumber \\
\delta_1 A &=& 2a_2 B^a \xi_a +\frac{b_2}{2} \varphi e_a \xi^a, \\
\delta_1 \varphi &=& \frac{1}{(d-1)(d-2)} [ 2a_3 (\pi\xi) - b_2
\varphi \xi]. \nonumber
\end{eqnarray}
To obtain these particular form of the vertex we used the following
field redefinitions (which do not raise the number of derivatives):
\begin{eqnarray}
f^a &\Rightarrow& f^a + \kappa_1 f^a \varphi + \kappa_2 e^a \varphi^2,
\nonumber \\
A &\Rightarrow& A + \kappa_3 \varphi A, \\
\varphi &\Rightarrow& \varphi + \kappa_4 \varphi^2. \nonumber
\end{eqnarray}
In the unitary gauge this vertex has a very simple form:
\begin{equation}
{\cal L}_1 = a_0 \hat{E}_{a[3]} [ \Omega^{ab} \Omega^a{}_b f^a +
\Omega^{a[2]} \Omega^{ab} f_b ] + c_1 \hat{E}_{a[3]} f^a f^a f^a.
\end{equation}

\subsection{Gravitational interaction}

In this case for the minimal cubic vertex we obtained:
\begin{eqnarray}
{\cal L}_1 &=& a_0 \hat{E}_{a[3]} [ \Omega^{ab} \Omega^a{}_b h^a +
\Omega^{a[2]} \Omega^{ab} h_b + \Omega^{a[2]} \omega^{ab} f_b ] 
+ \frac{a_0}{4} \hat{E}_{a[4]} R^{a[2]} f^a f^a \nonumber \\
 && + a_0 \hat{a}_a ] B_{b[2]} B^{b[2]} h^a + 4 B^{ab} B_{bc} h^c ]
- (d-1)(d-2)a_0 \hat{e}_a [ (\pi\pi) h^a - 2 \pi^a \pi^b h_b ]
\nonumber \\
 && + b_1 \hat{E}_{a[3]} \Omega^{a[2]} A h^a + b_2 \hat{E}_{a[2]}
B^{a[2]} f^b h_b \nonumber \\
 && + c_1 \hat{E}_{a[3]} f^a f^a h^a + c_2 \hat{E}_{a[2]} f^a h^a
\varphi + c_3 \hat{e}_a h^a \varphi^2, 
\end{eqnarray}
where
$$
b_1 = - \frac{2ma_0}{(d-2)}, \qquad
b_2 = - ma_0,
$$
$$
c_1 = - M^2a_0, \qquad
c_2 = (d-1)Mma_0, \qquad
c_3 = - d(d-1)m^2a_0.
$$
Here the corrections to the physical fields transformations have the
form:
\begin{eqnarray}
\delta_1 f^a &=& 2a_0 \hat{\eta}^{ab} f_b - 2a_0 \Omega^{ab} 
\hat{\xi}_b + 2a_0 \eta^{ab} h_b - 2a_0 \omega^{ab} \xi_b 
- 2b_1 A \hat{\xi}^a + 2b_1 h^a \xi, \nonumber \\
\delta_1 A &=& 2a_0 e_a B^{ab} \hat{\xi}_b - b_2 f^a \hat{\xi}_a
+ b_2 h^a \xi_a, \qquad \delta_1 \varphi = - 2a_0 (\pi \hat{\xi}), \\
\delta_1 h^a &=& 2a_0 \eta^{ab} f_b - 2a_0 \Omega^{ab} \xi_b 
+ \frac{4(d-3)ma_0}{(d-2)} [f^a \xi - A \xi^a ]. \nonumber
\end{eqnarray}
In this case we also used the allowed field redefinitions:
\begin{eqnarray}
f^a &\Rightarrow& f^a + \kappa_1 h^a \varphi, \nonumber \\
h^a &\Rightarrow& h^a + \kappa_2 F^a \varphi + \kappa_3 e^a
\varphi^2. 
\end{eqnarray}
In the unitary gauge this vertex looks as follows:
\begin{eqnarray}
{\cal L}_1 &=& a_0 \hat{E}_{a[3]} [ \Omega^{ab} \Omega^a{}_b h^a +
\Omega^{a[2]} \Omega^{ab} h_b + \Omega^{a[2]} \omega^{ab} f_b ]
\nonumber \\
 && + \frac{a_0}{4} \hat{E}_{a[4]} R^{a[2]} f^a f^a - M^2a_0 
\hat{E}_{a[3]} f^a f^a h^a.
\end{eqnarray}

\end{document}